\documentclass[aps,preprint,showpacs]{revtex4}
\usepackage{graphicx}

\begin{document}

\title{ The differential sum rule for the relaxation rate in 
 dirty superconductors 
}
\author{Andrey V. Chubukov$^1$, Ar. Abanov$^2$, and D.N. Basov$^3$}
\affiliation{$^1$Department of Physics, University of Wisconsin, Madison, WI 53706}
\affiliation{$^2$Los Alamos National Laboratory, MS 262B, Los Alamos, NM 87545}
\affiliation{$^3$ Department of Physics, University of California, San Diego, La Jolla,
CA 92093}
\date{\today}

\begin{abstract}
We consider the differential sum rule for the  effective scattering rate $%
1/\tau (\omega)$ and optical conductivity $\sigma _{1}(\omega ) $ in a 
dirty BCS superconductor, for arbitrary ratio of the superconducting gap $%
\Delta$ and the normal state constant damping rate $1/\tau$. We show that if 
$\tau$ is independent of $T$,  the area under $1/\tau (\omega)$ does not
change between the normal and the superconducting states, i.e., there exists
an exact  differential sum rule for the scattering rate.  For \textit{any}
value of the dimensionless parameter $\Delta\tau $, the sum rule is
exhausted at frequencies controlled by $\Delta$. 
%but the numerical convergence is weak. 
We show that in the dirty limit the convergence of the differential 
sum rule for the scattering rate is much faster then the convergence of 
the $f-$sum rule, but slower then the convergence of the differential 
sum rule for conductivity.
\end{abstract}

\pacs{71.10.Ca,74.20.Fg,74.25.-q}
\maketitle

%\twocolumn[\hsize\textwidth\columnwidth\hsize\csname @twocolumnfalse\endcsname   

%\address

%\address

%\draft   
%\maketitle    

%]
%\narrowtext 

Optical constants of solids follow a variety of sum rules.~\cite{pi_noz} The
origins of the sum rules can be traced back to fundamental conservation laws
and are intimately connected to the causality of the electromagnetic
response leading to Kramers-Kronig relationships between the real and
imaginary parts of the optical constants.  The analysis of the sum rules is
a powerful tool to study the distribution of the spectral weight in
correlated electron systems. 

The dynamics of conducting carriers
 is usually described in terms of the effective scattering rate 
$1/\tau (\omega)$ and the effective mass $m^* (\omega)$~\cite{tom} 
\begin{equation}
\tau ^{-1}(\omega )=\frac{\omega _{pl}^{2}}{4\pi }~\Re[\frac{1}{\sigma
(\omega )}],~\frac{m^*}{m_b} = -\frac{\omega _{pl}^{2}}{4\pi } \frac{1}{\omega}
~\Im[\frac{1}{\sigma
(\omega )}]
\end{equation} 
where $\sigma (\omega )=\sigma
_{1}(\omega )+i\sigma _{2}(\omega )$ is the complex optical conductivity, $%
\omega _{pl}^{2}=4\pi ne^{2}/m_{b}$ is the plasma frequency, and $%
m_{b}$ is the band mass~\cite{tom}. 
 Under special circumstances (e.g., when the Eliashberg theory is valid),
 the spectra of $1/\tau 
(\omega)$ and $m^*$ can be linked to the real and imaginary parts
 of the electronic self energy. 

The conductivity, that characterizes the absorbing power of a solid, obeys
the  $f-$sum rule $\int_{0}^{\infty }d\omega \sigma _{1}(\omega )=\omega
_{pl}^{2}/8$. This sum rule reflects the conservation of the number of
particles and physically implies that at a given particle density, the total
absorbing power does not depend on the details of the interactions, and is
determined only by the total number of particles in the system ~\cite{pi_noz}%
. It is particularly relevant for a superconductor where the conductivity
acquires a $\delta $-functional piece proportional to the superfluid density 
$n_{s}$, and the sum rule transforms into $\int_{+0}^{\infty }d\omega \sigma
_{1}(\omega )=\omega _{pl}^{2}/8(1-n_{s}/n)$. As $n_{s}/n$ can be measured
independently, the sum rule is a valuable tool to study the transformation
of the spectral weight associated with the superconductivity.

The issue raised in series of recent 
works~\cite{basov,marsiglio,ac_sum,homes}  is whether there exists any
similar sum rule for the relaxation rate. At the first glance, the answer is
negative as the  integral over $1/\tau (\omega)$ diverges and hence has no
physical meaning. 
This can be readily illustrated with a simple Drude model
 where  $%
\sigma (\omega) = (\omega_{pl}^{2}/4\pi )/(1/\tau - i \omega)$, i.e.,  $\tau
(\omega) = \tau$, and  $\int_0^{\infty} d \omega \tau^{-1} (\omega)$ is
infinite. This  implies that there is no sum rule  for $1/\tau (\omega)$
that could be traced to a conservation law.  However, one still can argue
that the frequency integral of the scattering rate has some physical
meaning.  Basov et al.~\cite{basov} argued, in connection with the cuprates,
 that for experimentally relevant
frequencies $1/\tau (\omega)$ is numerically close to  an effective
scattering rate $1/\tau_{sr}$ (defined below) that is related to the dielectric  function and obeys a physically motivated  sum rule. In this
situation, one can argue that there should exist an approximate sum rule  if
the frequency integral over $1/\tau (\omega)$ is taken in finite limits. 
This approach was thoroughly analyzed  by Marsiglio et al.~\cite{marsiglio}.

In this paper, we analyze whether there exists 
  an \textit{exact}
sum rule for the full frequency integral of the difference between $1/\tau (\omega)$ at two different temperatures in the normal state, and, 
what is actualy more relevant, between $1/tau (\omega)$
in the normal and superconducting states. 
  For a Drude metal, the frequency
integral over the difference of $1/\tau (\omega )$ at different $T$ 
in the normal state is
either infinite or vanishes depending on whether or not $1/\tau $ depends on 
$T$. We argue that zero and infinity are the only two options for the
differential integral in any model, 
and that if $\tau (\omega)$ is $T$ independent,
  the
differential sum rule for the relaxation rate is \textit{exact}  between
the normal state and a BCS superconductor.

We also analyze
 what controls the rate of convergence of the sum rules for conductivity and $1/\tau$. This issue is elevant for experimental applications as when 
 sum rules are applied to the analysis of the experimental data,
 one always encounters a problem that actual data is 
available over limited frequency intervals.  
It is therefore imperative to know rapidly sum rules converge
 to estimate the accuracy of the experiment-based sum rule analysis.
 
The analysis in the clean limit $\Delta \tau \gg 1$,  where $\Delta$ is the
superconducting gap, and  $\tau$ is  independent on $T$ was 
 performed earlier~\cite{ac_sum}, as a byproduct of the
 modeling for the cuprates.
 In this limit, $\Delta$ is the largest energy scale in the problem, and 
 the
differential sum rule between the normal state and a BCS superconductor 
is exact and is  exhausted at frequencies of order $\Delta$. 
Here we consider arbitrary $\Delta \tau$. We show that even in
the dirty limit $\Delta \tau \ll 1$, the differential sum rule is  exact 
 and is furthermore still 
exhausted at frequencies of order $\Delta$. 
This  result is not intuitively obvious as the 
conductivity sum rule in the dirty limit is 
exhausted only at $\omega \sim 1/\tau \gg \Delta$.
We also found that the  
  functional form of 
 $1/ \tau_{sc} (\omega)$ only weakly depends on 
$\Delta \tau$  at $\omega > 2 \Delta$. Here and below 
we use subscripts ``$sc$'' and ``$n$'' 
for superconducting and normal states, respectively

The fact that the differential sum rule for $1/\tau$ is exausted
at $\omega \propto \Delta$ in the dirty limit may be relevant
for the interpretation of the data from some high $T_c$ materials.
The frequency below which fermionic excitations become incoherent
may arise from the energy scale $\omega \sim 1/\tau$ .  The convergence
of the differential sum rule at $O(\Delta)$ and the near independence
of $\tau_{sc}(\omega)$ on $\Delta\tau$ implies that as long as the system
possesses a sharp superconducting gap the differential sum rule
for $1/\tau$ is not affected by the increased fermionic incoherence.
Indeed in high $T_c$ superconductors, $1/\tau$ is temperature dependent.
The issue of whether or not the differential sum rule is exact between
the normal and superconducting states could then only be experimentally
addressed if measurements could be done in both states at the
same T.  This is obviously impossible.   What can be clearly observed, 
at least in YBCO,
is the re-distribution of the spectral weight between 
the normal and superconducting states.  If this redistribution
is over scales of $O(\Delta)$ then the frequency integral can be
restricted to a few times $\Delta$.  The approximate sum rule can
then be analyzed in the hope that the T dependence of the relaxation rate
 is a minor effect
compared to the huge change in $1\tau(\omega)$ which takes place
between the normal and superconducting states
Note in this regard  that since $1/\tau$ is expressed via both $\Re \sigma 
(\omega)$ and $\Im \sigma (\omega)$, the analysis of the 
differential sum rule for $1/\tau$, even though it is only 
approximate, still yields information about the spectral weight 
distribution in a superconductor, which is complimentary to the 
information from the $f-$sum rule that involves only  $\Re \sigma (\omega)$  

The issue whether or not $1/\tau $ differential sum rule is satisfied in
dirty BCS superconductors was a subject of recent controversy. Basov et al. 
\cite{basov} argued that in both clean and dirty limits, $1/\tau
_{sc}(\omega )$ vanish below $2\Delta $, but overshoots the normal state $%
1/\tau $ at larger frequencies. In that paper, the profile of $1/\tau
_{sc}(\omega )$ was visually related to that of the fermionic density of
states, for which the sum rule is exact and reflects the conservation of the
number of particles. Homes et al. \cite{homes} correctly observed that the
relaxation rate and conductivity are expressed via a current-current
correlator, and therefore scale with the joint density of states of two
fermions about which, they conjectured, no rigorous statements can be made.
They computed the frequency integral of $1/\tau_{sc}(\omega )-1/\tau
_{n}(\omega )$ numerically for a BCS superconductor $\tau_{n}(\omega )=\tau 
$ with $\tau \Delta =1/2$ and found that even at $\omega $ as high as $%
15\Delta $, it is still about $25\%$ of its maximum value at $\omega
=2\Delta $. They concluded, based on this numerics, that the value of sum
rule integral depends on the choice for cutoff frequency, and that if the
integral is truncated at $\omega =O(\Delta )$, a finite value is expected.
Below we demonstrate explicitly that the differential sum rule is in fact
exhausted at the energy scale controlled by $\Delta $, however, a rather slow
convergence found in \cite{homes} is real and is due to a weak $(\log \omega
)/\omega ^{2}$ decay of $1/\tau_{sc}(\omega )$ at large frequencies.

We begin with the general argument that  zero and infinity are the only two
options for $\int_0^\infty d \omega/\tau (\omega)$. 
This conjecture can be verified by applying
the Kubo formula that  relates $\sigma (\omega)$ with the full retarded
current-current correlator $\Pi (\omega)$: $\sigma (\omega) =
(\omega^2_{pl}/4\pi)~\Pi (\omega)/(-i\omega + \delta)$. Substituting this
relation into $1/\tau (\omega)$ we find that  $1/\tau (\omega) = \Im
S(\omega)/\omega$ where $S(\omega) = - \omega^2/\Pi (\omega)$. Both $\Im \Pi
(\omega)$ and $S(\omega)$  are odd functions of frequency. Since $\Pi
(\omega)$ is analytic in the upper half-plane of complex $\omega$  and does
not have zeros (that can be checked explicitly, see below), $S(\omega)$ is
also an analytic function in the upper half-plane. The analyticity implies
that, by Kramers-Kronig relations, 
\begin{equation}
\int_0^\infty \frac{1}{\tau (\omega)} = \int_0^\infty \frac{S (\omega)}{%
\omega} d \omega = \frac{\pi}{2}~\Re S (0) + C  \label{1a}
\end{equation}
where $C=0$ if the integral converges, and $C = \infty$ if it is not. It is
easy to show that $\Re S(0)$ vanishes  both in the normal and in the
superconducting state. In the normal state, $\Re \Pi (\omega) \propto
\omega^2,~\Im \Pi (\omega) \propto \omega$ at the lowest frequencies, hence $%
S(\omega) \propto \omega^2$ and $S(0) =0$. In the superconducting state $\Re
\Pi (0)$ is finite, while $\Im \Pi (0) =0$, hence $S(\omega)$ again scales
as $\omega^2$, and again $S(0) =0$. Hence $\int d \omega/\tau (\omega) =C$,
i.e., it is either zero or infinite. For non-differential sum rule, $C =
\infty$ as at  large frequencies, 
%fermions behave as almost free quasiparticles, 
$\Pi (\omega)$ tends to 1, hence $S(\omega) \approx \omega^2$,
 and the integral in (\ref{1a}) diverges. This confirms that there is no
physically motivated sum rule for $1/\tau (\omega)$.

There are two ways to improve the situation. First, one can introduce  an
effective $S_{eff}(\omega)$ that converges at 
high frequencies and is close to  $%
S(\omega)$ at experimentally relevant frequencies. Then one can hope to
obtain an experimentally meaningful approximate  sum rule. For the
scattering rate, the natural choice, suggested by Basov et al.~\cite{basov}
is to introduce an effective scattering rate 
\begin{equation}
\frac{1}{\tau_{eff}} = \frac{\omega^2_{pl}}{\omega} \Im \left[1 - \frac{1}{%
\epsilon (\omega)}\right]  \label{1b}
\end{equation}
where $\epsilon (\omega) = 1 + 4 \pi i \sigma (\omega)/\omega$ is the
dielectric function. For this scattering rate, $S_{eff} (\omega) = 1 -
\omega/(\omega + 4 \pi i \sigma (\omega))$ is an analytic function, and at
high frequencies it vanishes as $S_{eff} (\omega) = O(1/\omega^2)$. The
Kramers-Kronig relations are then applicable, and 
\begin{equation}
\int_0^\infty \frac{d \omega}{\tau_{eff} (\omega)} = \frac{\pi \omega^2_{pl}}{%
2} \Re \left[1-\frac{1}{\epsilon(\omega ) }\right]_{\omega =0} = \frac{\pi}{2}%
\omega _{pl}^{2}  \label{2a}
\end{equation}
For a Drude metal, 
\begin{equation}
\frac{1}{\tau_{eff} (\omega)} = \frac{1}{\tau } \frac{\omega_{pl}^{4}}{%
(\omega^{2}-\omega_{pl}^{2})^{2}+\omega^{2}/\tau^{2}}  \label{3a}
\end{equation}
At frequencies $\omega \ll \omega_{pl}$, the correction term in the
denominator can be neglected, and $1/\tau_{eff} (\omega) \approx 1/\tau$. As
the plasma frequency can well be larger than the fermionic bandwidth,
the actual integration of experimentally measured $1/\tau (\omega)$ may not
extend to $\omega \sim \omega_{pl}$, i.e., in the measured range 
$1/\tau_{eff} (\omega)$ and $%
1/\tau (\omega)$ are nearly the same. Still, however, this does not imply that
 there is an ``approximate'' sum rule for $1/\tau$ as 
$\int d \omega/\tau (\omega)$ and $\int d
\omega/\tau_{eff} (\omega)$ both diverge when the integration is restricted 
to $\omega \ll \omega_{pl}$. Only when the frequency integration extends to 
$\omega > \omega_{pl}$, $\int_0^\omega dx/\tau_{eff} (x)$ converges to the
 sum rule value. In other words, typical frequencies for 
$\int_0^\infty 1/\tau_{eff} (\omega)$ are of order $\omega_{pl}$, and at 
these frequencies $1/\tau$ and $1/\tau_{eff}$ are very different.

As we pointed out above, in this paper we use another approach and analyze the 
difference between $1/\tau (\omega)$ in the normal and superconducting
states. The leading divergent term in $S(\omega) \sim \omega^2$ at high
frequencies is the same for both states. It therefore cancels out in $S_{sc}
(\omega) - S_n (\omega)$. If the subleading terms scale as negative powers
of frequency at  large $\omega$, $S_{sc} (\omega) - S_n (\omega)$ converges
and then the sum rule becomes exact. We emphasize that contrary to
$S_{eff} (\omega)$, the difference  of $S_{sc} (\omega) - S_n (\omega)$, 
if converges, begins
falling off at frequencies that are still much smaller than the fermionic
bandwidth and $\omega_{pl}$. In other words, the differential sum
rule for the scattering rate is exact in a continuous model.

We now check whether the differential sum rule for $1/\tau $ is satisfied in
a BCS superconductor (we argue that it is exact) and also examine the energy
scale at which this sum rule  is exhausted. In order to explore the issue of
convergence we will analyze the differential frequency sums for both $1/\tau
(\omega )$ and $\sigma _{1}(\omega )$ defined as 
\begin{eqnarray}
I_{\tau }(\omega ) &=&\int_{0}^{\omega }d\Omega (\tau_{sc} ^{-1}(\Omega )-\tau_n^{-1})  \label{1} \\
 I_{\sigma }(\omega ) &=&\int_{0}^{\omega }d\Omega (\sigma _{1,sc}(\Omega
)-\sigma _{1,n}(\Omega ))  \label{1s}
\end{eqnarray}
where $\tau_n = \tau$ and, as we pointed out above, the notations ``sc'' and ``n'' refer to superconducting and normal states, respectively.
 
The expression for the current-current polarization operator  in a dirty BCS
superconductor has the form 
\begin{eqnarray}
&&\Pi_{sc} (\omega) =\int_0^{\infty} d \Omega ~\frac{1}{\left(\sqrt{%
\Omega^2_+ - \Delta^2} + \sqrt{\Omega^2_- - \Delta^2} + i/\tau\right)} 
\nonumber \\
&&\times \frac{\sqrt{\Omega^2_+ - \Delta^2} \sqrt{\Omega^2_- - \Delta^2} -
\Delta^2 - \Omega_+ \Omega_-}{\sqrt{\Omega^2_+ - \Delta^2} \sqrt{\Omega^2_-
- \Delta^2} }   \label{3}
\end{eqnarray}
%\end{widetext}
where $\Omega_{\pm} = \Omega \pm \omega/2$. In the normal state, this
reduces to a conventional Drude form $\Pi_n (\omega) = \omega/(\omega +
i/\tau)$. In the superconducting state, one can show quite generally that $%
\Im \Pi (\omega)$ vanishes below $2 \Delta$. At high frequencies, $\Pi
(\omega)$  gradually approaches $\Pi (\infty) =1$.

We first verify whether $S_{sc}(\omega )-S_{n}(\omega )$ vanishes at $\omega
=\infty $. As typical internal frequencies in Eq. (\ref{3}) are of the same
order as external $\omega $, at $\omega \gg \Delta $, one can expand the
integrand in (\ref{3}) in $\Delta /\omega $. The expansion can be
straightforwardly carried out for arbitrary $\Delta \tau $, and the result
is that at high frequencies 
\begin{eqnarray}
\Re\Pi _{sc}(\omega )\!\!\! &=&\!\!\!\Re\Pi _{n}(\omega )\left( 1-\frac{1}{%
1+(\omega \tau )^{2}}\frac{2\Delta ^{2}\log \frac{\omega }{\Delta }}{\omega
^{2}}\right)   \nonumber \\
\Im\Pi _{sc}(\omega )\!\!\! &=&\!\!\!\Im\Pi _{n}(\omega )\left( 1+\frac{%
(\omega \tau )^{2}-1}{(\omega \tau )^{2}+1}~\frac{2\Delta ^{2}\log \frac{%
\omega }{\Delta }}{\omega ^{2}}\right)   \label{5}
\end{eqnarray}%
where $\Re\Pi _{n}(\omega )=(\omega \tau )^{2}/(1+(\omega \tau )^{2}),~\Im\Pi
_{n}(\omega )=-\omega \tau /(1+(\omega \tau )^{2})$. Substituting these
results into $S(\omega )=\omega ^{2}/\Pi (\omega )$ we find that for large $%
\omega \gg 1/\tau $~ 
\begin{equation}
\Re[S_{sc}(\omega )-S_{n}(\omega )]\propto \omega ^{-2}~~
\Im[S_{sc}(\omega )-S_{n}(\omega )]\propto \omega ^{-1}.
\end{equation}
 We see that
both real and imaginary parts of $S_{sc}(\omega )-S_{n}(\omega )$ vanish at
infinite frequency. This implies that Kramers-Kronig transformation is
applicable, and hence \textit{the differential sum rule is exact for a dirty
BCS superconductor}.
% provideed that the normal state quasiparticle lifetime
% $\tau$ is $T$ independent. 

\begin{figure}[tbp]
%\includegraphics[scale=1.2]{WZNW}
%\begin{figure}[tbp]
%\epsfxsize= 3.2in
%\begin{center}
%\leavevmode
%\epsffile{cleanSigmaTau.eps}
%\end{center}
%\vspace{-0.75cm}
%
\includegraphics[clip=true,width=0.8\columnwidth]{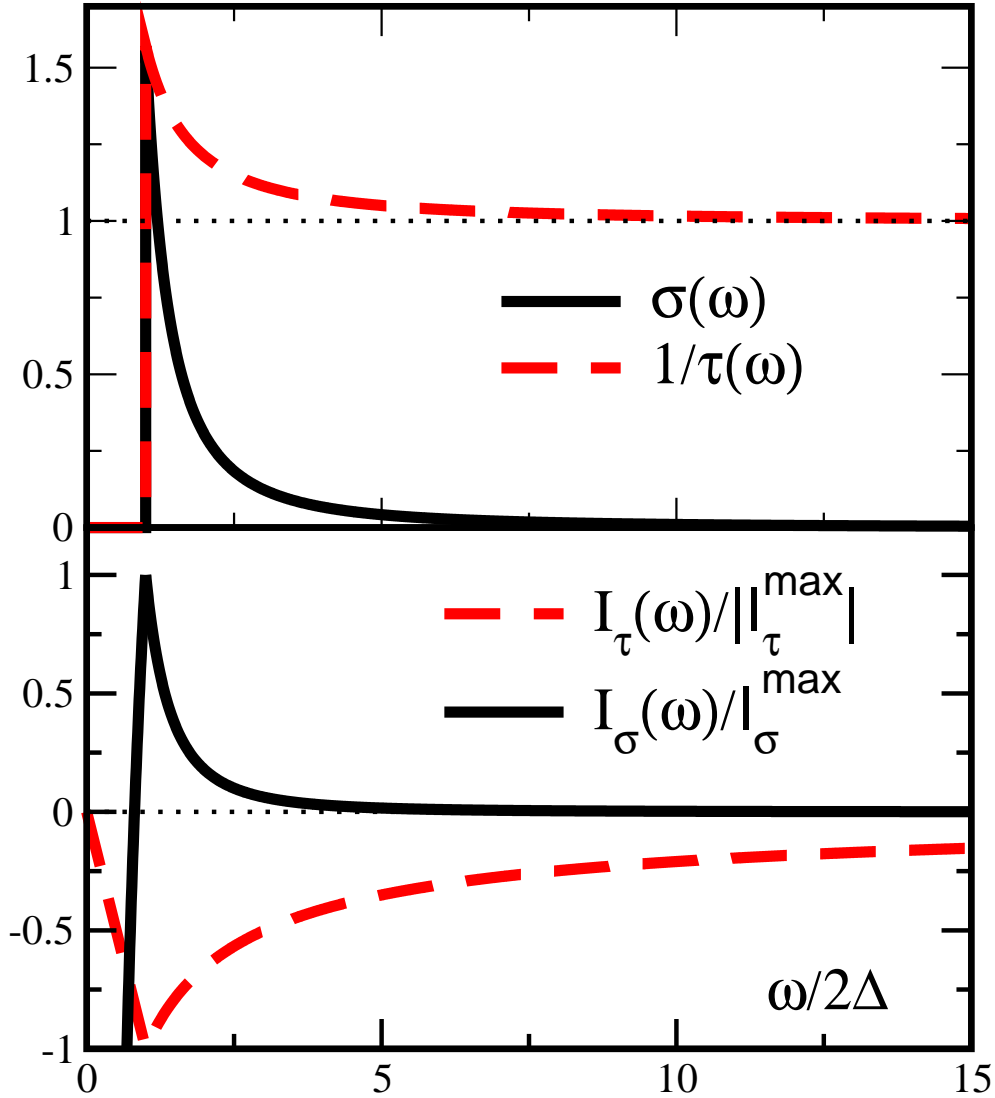} 
\caption{(a) The frequency dependence of the relaxation rate $\protect\tau /%
\protect\tau_{sc} (\protect\omega)$ in 
%a BCS superconductor in 
the clean limit. (b) The behavior of $I_{\protect\tau} (\protect\omega)$
  and $I_{\protect\sigma} (\protect\omega)$ normalized to their maximum values.}
\label{fig1}
\end{figure}

We now address the issue of the energy scale 
over which the sum rule is exhausted. We 
 consider clean and dirty limits separately. 
 In the clean limit,  the
frequency integral in (\ref{3}) was evaluated in Ref.~\cite{ac_sum}.  To
first order in $1/\Delta \tau$ we have for $\omega > 2\Delta$
% where $\Im\Pi_{sc} (\omega)$ and hence $1/\tau_{sc} (\omega)$ are finite 
\begin{equation}
\Re \Pi_{sc} (\omega) \approx 1,~\Im \Pi_{sc} (\omega) \approx -\frac{1}{\tau
\omega} E\left(\sqrt{1 - \frac{4\Delta^2}{\omega^2}}\right),  \label{8}
\end{equation}
where $E(x) = \int_0^{\pi/2} d \phi \sqrt{1 - x^2 \sin^2 \phi}$ is the
complete Elliptic integral of the second kind (note that definitions of $E$
differ in different handbooks). In the two limits $E(0) = \pi/2$ and $E(1) =1$.
The result for $\Re \Pi_{sc} (\omega) \approx 1$ is valid outside a tiny $%
O(1/\Delta \tau)$ range near $2\Delta$ where $\Re \Pi_{sc} (\omega)$ 
diverges logarithmically. Substituting $\Pi_{sc} (\omega)$ from (\ref{8}) 
into $1/\tau_{sc} (\omega)$ we obtain for $\omega >2\Delta $ 
\begin{equation}
\frac{1}{\tau_{sc} (\omega)} = \frac{1}{\tau} E\left(\sqrt{1 - \frac{%
4\Delta^2}{\omega^2}}\right),~~
 \sigma_{1} (\omega ) = \frac{\omega _{pl}^{2}%
}{4\pi }\frac{1}{\omega ^{2}} \frac{1}{\tau_{sc} (\omega)}  \label{9s}
\end{equation}
and $1/\tau (\omega ) = \sigma (\omega ) =0$ for $\omega <2\Delta $. We plot
these functions in Fig. \ref{fig1}. For the differential sum rule we then
obtain ($\omega >2\Delta $) 
\begin{eqnarray}
\!\!\!&&I_{\tau} (\omega) = \frac{2 \Delta}{\tau}~\int_0^{\frac{\omega}{%
2\Delta}} dx \left[-1 + \Re E\left(\sqrt{1 - \frac{1}{x^2}}\right)\right]
\label{10} \\
\!\!\!&&I_{\sigma }(\omega ) \!=\! \frac{\omega ^{2}_{pl}}{4\pi }\!\frac{1}{%
2\Delta \tau } \left[\frac{2\Delta }{\omega }\!-\!\frac{\pi ^{2}}{8}%
\!+\!\int_{\frac{2\Delta}{\omega} }^{1}\!\!\!dxE\left(\sqrt{1\!-\!x^{2}}
\right) \right]  \label{10s}
\end{eqnarray}
We explicitly verified that $I_{\tau} (\infty) =0$, i.e., the differential
sum rule is indeed satisfied. 

%
% It is instructive to verify the 
% differential sum rule  (i.e., that $I(\infty) =0$)  directly, without appealing to Kramers-Kronig relation.
% Substituting $E(z)$ into (\ref{10})
% and changing variables we obtain 
%\begin{eqnarray}
%&&I_{\tau} (\infty) = \frac{2 \Delta}{\tau} \left[-1 + \int_1^{\infty} dx \left(-1 + \Re E\left(\sqrt{1 - \frac{1}{x^2}}\right)\right)\right]\nonumber \\
%&& = \frac{2 \Delta}{\tau} \left[-1 + \int_0^1 dy \int_1^{\infty} dx \left(\frac{1}{x} \sqrt{\frac{x^2 (1-y^2) + y^2}{1-y^2}} -1\right)\right]
%\label{11}
%\end{eqnarray}
%Integrating  over $x$ first we obtain 
%\begin{equation}
%I_{\tau} (\infty) = \frac{\Delta}{\tau} \int_0^1 dy \left[\frac{y}{\sqrt{1-y^2}}~\log{\frac{1+y}{1-y}} - \frac{2}{\sqrt{1-y^2}}\right]
%\label{12} 
%\end{equation}
%Evaluating now the first integral in the r.h.s of (\ref{12}) by parts
% we find that it cancells the second integral, i.e., $I_{\tau} (\infty) =0$. 

We also see from (\ref{10}) that at finite $\omega$, $I_{\tau} (\omega)$
depends only on $\omega/\Delta$.  This implies that  in the clean limit, the
differential sum rule is exhausted at frequencies $O(\Delta)$. 
 An important issue, however,
 is  how rapidly $I(\omega)$ converges to zero at $\omega \gg \Delta$.
In Fig. \ref{fig1}b we plot $I_{\tau} (\omega)$ and $I_{\sigma }(\omega )$
evaluated numerically from (\ref{10}) and normalized to their values at the
maximum at $\omega = 2\Delta$. We see that $I_{\tau} (\omega)$ converges
much more slowly that $I_\sigma (\omega)$. In particular, at $\omega = 15
\Delta$, $|I_\tau (\omega)|$ is still about 25\% of its maximum
 value in accord with Ref.~\cite{homes}. 
On the contrary $I_s(\omega)$ is vanishingly small at $\omega=15\Delta$.

  value. This
result fully agrees with Ref.~\cite{homes}.  For the same $\omega$, $%
I_\sigma (\omega)$ practically vanishes.

A weaker convergence of $I_{\tau} (\omega)$ can be understood
analytically. Indeed, at high frequencies, the elliptic function can be
expanded  in $1/\omega^2$. This yields $1/\tau_{sc} (\omega) = (1/\tau)( 1 +
2 (\Delta/\omega)^2 \log (2\omega/\Delta))$. Integrating this expression
over frequency, we obtain  that at $\omega \gg \Delta$ 
\begin{equation}
I_{\tau} (\omega) = \frac{2 \Delta}{\tau}~\frac{\Delta}{\omega} \left[1/2 +
\log(2 \omega/\Delta)\right]  \label{14}
\end{equation}
The conductivity integral meanwhile converges to zero as  
\begin{equation}
I_\sigma (\omega) = \frac{\omega _{pl}^{2}}{4\pi } \frac{1}{12\Delta \tau }%
\left(\frac{2\Delta }{\omega } \right)^{3} \left[1/6+\ln (2\omega/\Delta )%
\right]  \label{15}
\end{equation}
Comparing (\ref{14}) and (\ref{15}), we see that $I_\sigma (\omega)$ has an
extra  $(2\Delta /\omega)^{2} $ that accounts for much faster convergence 
of $I_\sigma (\omega)$ than of $I_\tau (\omega)$. 
%  This logarithmical factor actually strongly influence numerical values.

We next proceed to the dirty limit $\Delta \tau \ll 1$. As we said in the
introduction, the key issue here is whether the sum rule is still exhausted
at frequencies $O(\Delta)$, or one needs to extend the integration to
frequencies of order $1/\tau$, where the $f-$sum rule for the conductivity
is exhausted. At the first glance, in the dirty limit, the frequency
integration has to be extended to larger frequencies than in the clean
limit, as one can easily show that at $\Delta \tau \ll 1$, the jump of $%
1/\tau_{sc} (\omega)$ at $2\Delta$ is small, of order $\Delta \tau$, and
therefore $1/\tau_{sc} (\omega)$ in a superconductor does not overshoot the
normal state $1/\tau$ immediately above $2\Delta$. However, as we now
demonstrate, typical frequencies for the differential sum rule still scale
with $\Delta$.

\begin{figure}[tbp]
\includegraphics[width=0.8\columnwidth]{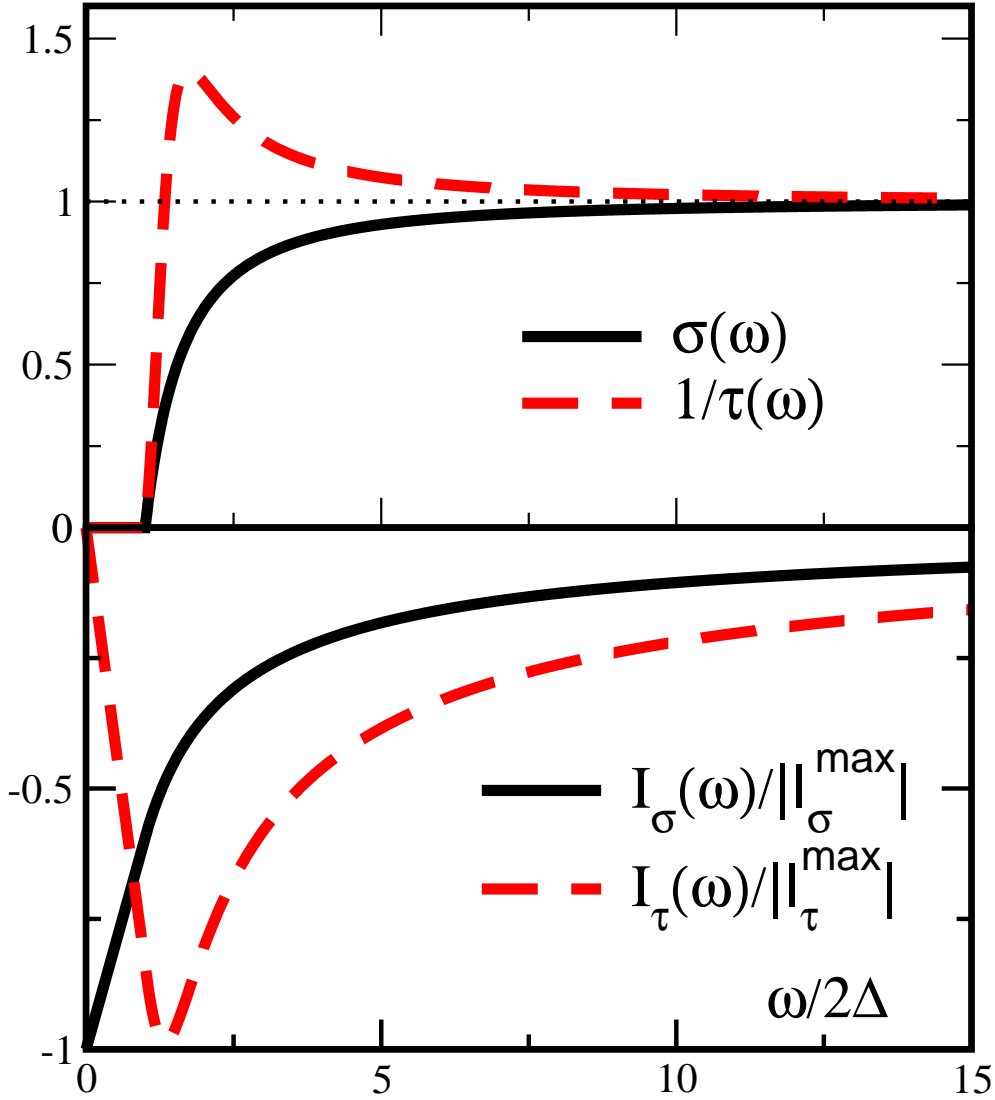}
\caption{Same as in Fig~\protect\ref{fig1} but in the dirty limit $\Delta 
\protect\tau \ll 1$. }
\label{fig2}
\end{figure}
%(c) The comparative behavior of 
% $(\tau/2\Delta) I_{\tau} (\omega)$ and 
%$I_{\sigma} (\omega)/I_{\sigma} (\infty)$ in clean and dirty limits.

The first indication that the physics is still confined to frequencies $%
O(\Delta)$ comes from the analysis of the form of $1/\tau_{sc} (\omega)$ at $%
\omega \sim \tau^{-1} \gg \Delta$. Using Eqs. (\ref{5}) for $\Pi (\omega)$
and substituting them into $1/\tau_{sc} (\omega)$, we obtain 
%after elementary manipulations 
\begin{equation}
\frac{1}{\tau_{sc} (\omega)} = \frac{1}{\tau} \left(1 + \frac{2 \Delta^2}{%
\omega^2} \log{\frac{2 \omega}{\Delta}} \frac{1 + (\omega \tau)^4}{(1 +
\omega \tau)^4}\right)  \label{16}
\end{equation}
We see that at frequencies comparable to $1/\tau$, $1/\tau_{sc} (\omega)$
exceeds $1/\tau$, i.e., the overshoot occurs at a lower frequency. Further,
integrating $1/\tau_{sc} (\omega)$ from $O(1/\tau)$ to infinity  we find
that the contribution to $I_{\tau} (\omega)$ from these frequencies is of
order $\Delta^2 |\log (\Delta \tau)|$. Meanwhile, the loss of $I_{\tau}$  in
the superconducting state between $0$ and $2 \Delta$ is $2 \Delta/\tau$ 
that in the dirty limit is much larger that $\Delta^2 |\log (\Delta \tau)|$.
This implies that even in the dirty limit, the dominant contribution to the
sum rule comes from frequencies well below $1/\tau$. At these frequencies,
the current-current polarization operator  can be evaluated exactly to
leading order in $\Delta \tau$ as one can pull $1/\tau$ out of denominator
of  the integral in the r.h.s. of (\ref{3}). The remaining integral is 
evaluated easily, and on substituting the result into $1/\tau_{sc} (\omega)$
we obtain for $\omega >2\Delta $ 
\begin{eqnarray}
\frac{1}{\tau_{sc} (\omega)} \!\!\!&=&\!\!\! \frac{1}{\tau} \frac{\omega}{%
\sqrt{\omega^2 - 4 \Delta^2}} \Re\left[E^{-1}\left(\frac{\omega}{\sqrt{%
\omega^2 - 4 \Delta^2}}\right)\right]  \nonumber  \\
\sigma _{1}(\omega ) \!\!\!&=&\!\!\! \frac{\omega _{pl}^{2}}{4\pi }\tau 
\sqrt{1-\frac{\omega ^{2}}{4\Delta ^{2}}} \Re\left[E\left(\frac{\omega}{\sqrt{%
\omega^2 - 4 \Delta^2}}\right)\right]  
\label{17}
\vspace{0.1cm}
\end{eqnarray}
and $1/\tau_{sc} (\omega ) = \sigma _{1}(\omega ) = 0$ for $\omega <2\Delta $%
. We plot these functions in Fig.\ref{fig2}.  The behavior of $1/\tau_{sc}
(\omega)$ near $2\Delta$ and at high frequencies can be well understood
analytically.  Near $2\Delta$, expanding $E(x)$ for large value of the
argument, we immediately obtain that  $1/\tau_{sc} (\omega)$ evolves
continuously  (up to corrections $O(\Delta \tau)$), and  very near $2\Delta$
behaves as $1/\tau_{sc} (\omega) = (1/\tau) [\omega^2 - 4 \Delta^2]/\omega^2$%
. Still, it overshoots $1/\tau$ at $\omega \sim 2.68 \Delta$ and develops a
maximum at $\omega = 3.48\Delta$. At larger frequencies, $1/\tau_{sc}
(\omega)$ approaches $1/\tau$ as  $1/\tau_{sc} (\omega) = (1/\tau)( 1 + 2
(\Delta/\omega)^2 \log (2\omega/\Delta))$, i.e., exactly the same way as in
the clean limit. Analyzing the frequency integrals in $I_{\tau} (\omega)$
and $I_{\sigma} (\omega)$ 
%in the same way as before
 we immediately make sure
that they converge at $\omega$ comparable to $2\Delta$.

%It is instructive to check that the differential sum rule is satisfied. 
%Separating the frequency integral in $I_{\tau} (\omega)$
% into integrals from $0$ to $2\Delta$ and over larger frequencies, we obtain, after rescaling variables
%\begin{eqnarray}
%&&I_{\tau} (\omega) = \frac{2\Delta}{\tau} \Big\{-1 + \nonumber \\
%&&\int_1^{\omega/(2\Delta)} 
%dx \left[-1 + \frac{ x}{\sqrt{x^2-1}}~\Re E^{-1}\left(\frac{x}{\sqrt{x^2-1}}\right)\right]\Big\} \nonumber \\
%&&I_{\sigma} (\omega) = 1
%\label{18}
%\end{eqnarray}
%where $x = \omega/2\Delta$. We see that the integrals converge  
%at $x = O(1)$, i.e.,   the compensation of the loss of 
%$I_{\tau}$ at $\omega <2\Delta$ comes from frequencies $\omega \geq 2\Delta \ll 1/\tau$. The  contributions from 
%frequencies comparable to $1/\tau$ only account
% for small corrections of order 
%$(\Delta \tau) \log (\Delta \tau)$ that we neglected in (\ref{18}).

In Fig. \ref{fig2}b we compare the rates of convergence  of $I_{\tau}
(\omega)$ and $I_\sigma (\omega)$ in the dirty limit. We see that, as in the
clean limit,  $I_{\sigma} (\omega)$ converges better. This time, it is 
related to the presence of the extra logarithmical term in the high
frequency expansion of $I_{\tau} (\omega)$. 
% Finally, in Fig.\ref{fig2}c we compare the rate of convergences of 
%$I_{\tau} (\omega)$ and $I_{\sigma} (\omega)$ 
%in the clean and dirty limits. 
We also see that  the rate of convergence is almost the same  in both clean
and dirty limits.  We recall that this result is not  obvious as in the
dirty limit, $1/\tau_{sc} (\omega)$ and $\sigma_1 (\omega)$ gradually
increases above at $2\Delta$, while in the clean limit, they both  jump at $%
2\Delta$ and immediately overshoot the normal state values  of $1/\tau$ and $%
\sigma_1$, respectively.

%Finally, it is instuctive to verify explicitly 
%that the differential sum rule for $1/\tau$ is satisfied.  
%Integrating in (\ref{18}) by parts using 
%$d E(k)/d k = (E(k) - K(k))/k^2$, 
%%~\cite{gradstein},
% where $K(k)$ is the elliptic integral of the first kind, 
% we can rewrite $I_{\tau} (\infty)$  as
%\begin{equation}
%I_{\tau} (\infty) = - \frac{2\Delta}{\tau} \Im \int_0^\infty \frac{z K(z)}{E^2(z)} dz
%\label{19}
%\end{equation}
%We couldn't evaluate this integral analytically, but numerical integration
% yields  $I_{\tau} (\infty) =0$
% with a very high accuracy.  

To conclude, in this paper we analyzed the differential sum rule for the
scattering rate and optical conductivity in a dirty BCS superconductor. We
demonstrated that this sum rule is exact if the normal state $\tau$ is
independent on temperature. For arbitrary  $\Delta \tau$, the sum rule is
exhausted at frequencies controlled by $\Delta$, but the convergence is
rather weak due to logarithmical terms. 
We showed that in the dirty limit the convergence of the differential 
sum rule for the  scattering rate is much faster then the convergence of 
the $f-$sum rule for the conductivity, but slower than for 
the differential sum rule for
conductivity. The latter  has the fastest
 convergence in both clean and dirty limits.

\begin{acknowledgments}
We acknowledge useful discussions with G. Blumberg, C. Homes, D. van der
Marel, F. Marsiglio, M. Strongin, and J. Tu. The research was supported by
NSF DMR (A. Ch.) , NSF and DOE (D.N.B) and by DR Project 200153 at
Los Alamos National Laboratory (Ar. A.).
\end{acknowledgments}

\end{document}